\documentclass[compsoc,conference,a4paper,10pt]{IEEEtran}

\usepackage{cite}
\usepackage[hidelinks]{hyperref}

\usepackage{booktabs}
\usepackage[export]{adjustbox}
\usepackage{xspace}
\usepackage{makecell}
\usepackage{multirow}
\usepackage{wasysym} 
\usepackage{pifont} 

\usepackage{url}

\usepackage{listings}
\usepackage{xcolor}
\definecolor{codeBackground}{RGB}{255,255,255}
\definecolor{codeComment}{RGB}{0,156,0}
\definecolor{codeKeyword}{RGB}{0,0,255}
\definecolor{codeLiteral}{RGB}{233,0,0}
\definecolor{codeString}{RGB}{233,0,0}
\definecolor{codeDefault}{RGB}{0,0,0}
\definecolor{codeNumbers}{RGB}{150,150,150}
\newcommand\eg{\emph{e.g.},\xspace}
\newcommand\ie{\emph{i.e.},\xspace}
\providecommand{\etal}{\emph{et al.}\xspace}
\newcommand{\crbug}[1]{\url{https://crbug.com/#1}}
\newcommand{\crrev}[1]{\url{https://crrev.com/c/#1}}

\renewcommand{\texttt}[1]{%
 \begingroup
 \ttfamily
 \begingroup\lccode`~=`/\lowercase{\endgroup\def~}{/\discretionary{}{}{}}%
 \begingroup\lccode`~=`[\lowercase{\endgroup\def~}{[\discretionary{}{}{}}%
 \begingroup\lccode`~=`.\lowercase{\endgroup\def~}{.\discretionary{}{}{}}%
 \catcode`/=\active\catcode`[=\active\catcode`.=\active
 \scantokens{#1\noexpand}%
 \endgroup
}

\newif\ifstatus
\statusfalse

\begin{document}

\title{\textit{Chrowned} by an Extension: Abusing the Chrome DevTools Protocol through~the~Debugger~API}

\author{\IEEEauthorblockN{José Miguel Moreno}
\IEEEauthorblockA{\textit{Universidad Carlos III de Madrid} \\
Madrid, Spain \\
josemore@pa.uc3m.es}
\and
\IEEEauthorblockN{Narseo Vallina-Rodriguez}
\IEEEauthorblockA{\textit{IMDEA Networks Institute} \\
Leganés, Spain \\
narseo.vallina@imdea.org}
\and
\IEEEauthorblockN{Juan Tapiador}
\IEEEauthorblockA{\textit{Universidad Carlos III de Madrid} \\
Madrid, Spain \\
jestevez@inf.uc3m.es}
}

\maketitle

\begin{abstract}
  The Chromium open-source project has become a fundamental piece of the Web
  as we know it today, with multiple vendors offering browsers based on its
  codebase.
  One of its most popular features is the possibility of altering or
  enhancing the browser functionality through third-party programs known as browser
  extensions. Extensions have access
  to a wide range of capabilities through the use of APIs exposed by Chromium.
  The Debugger API---arguably the most powerful of such APIs---allows extensions to
  use the Chrome DevTools Protocol (CDP), a capability-rich tool for debugging and
  instrumenting the browser.
  In this paper, we describe several vulnerabilities present in the
  Debugger API and in the granting of capabilities to extensions
  that can be used by an attacker to take control of the
  browser, escalate privileges, and break context isolation. We demonstrate
  their impact by introducing six attacks that allow an attacker to
  steal user information, monitor network traffic, modify site
  permissions (\eg access to camera or microphone), bypass security
  interstitials without user intervention, and change the browser settings.
  Our attacks work in all major Chromium-based browsers as they are rooted
  at the core of the Chromium project.
  We reported our findings to the Chromium Development Team, who already
  fixed some of them and are currently working on fixing the remaining ones.
  We conclude by discussing how questionable design decisions, lack of public
  specifications, and an overpowered Debugger API have contributed to enabling
  these attacks, and propose mitigations.
\end{abstract}

\section{Introduction}
\label{sec:introduction}

Modern browsers expose powerful
APIs~\cite{google-api-reference,
google-permissions} to enable the development of third-party browser
extensions~\cite{mdn-browser-extensions, google-what-are-extensions}.
These are small programs, built and released by third-party developers, that
extend or enhance the default features offered by the browser like
cookie management and ad-blocking.
Because of their potential for abuse, these APIs
are permission-protected and not all of them
are automatically granted by default to extensions~\cite{google-permissions}.
In the case of Chromium, its permission system is quite limited and does not offer
a comprehensive set of protections as implemented, for example, in
Android~\cite{android-permissions}.
Anecdotal evidence has already shown how
these APIs have been abused for nefarious purposes~\cite{malwarebytes-extensions-that-lie,
zdnet-extensions-all-urls}.

One overlooked Chromium feature is the \texttt{debugger}
permission~\cite{google-debugger}, which grants access to a limited version of
the powerful Chrome DevTools Protocol (CDP), a core Chromium component for
debugging and instrumenting the browser through a command passing interface.
CDP is widely used for running End-to-End (E2E)
tests on web-based applications through popular tools like
Selenium, Puppeteer and Playwright, and for building crawlers.
CDP exposes a WebSocket server to which external applications can connect
to. Chromium extensions may also communicate with this component
using the Debugger API, which is protected by the
\texttt{debugger} permission.
The Debugger API is a general substitute
of virtually any other extension API as it
grants total control over tabs, windows and
critical browser resources. These powerful capabilities are expected to be
found in a debugging tool, but are also an obvious candidate for abuse if
they are insecurely exposed to potentially malicious actors.

Despite the risks of granting third-party
extensions access to such a powerful component,
no previous work has systematically analyzed the robustness of the
Debugger API implementation and its security implications.
In fact, Chromium's Debugger API is already being used by
at least 434 extensions published on the Chrome Web
Store according to a permission measurement that we performed
in June 2022.
Furthermore, no official specification detailing the
design and purposes of this component can be publicly found.
In this paper, we describe the results of a systematic security analysis done over the
Debugger API and related components in the Chromium codebase.
Our analysis focuses on finding violations of a set of
security requirements that we derive from Chromium's CRX API Security
Checklist~\cite{chromium-security-checklist}.
Through a systematic code review,
we find multiple vulnerabilities that can be exploited by a
third-party extension to $(i)$ circumvent
the permission model to elevate privileges and gain control over more
capabilities than expected, including key browser features;
and $(ii)$ break the isolation principles implemented by Chromium
to prevent an attacker from accessing third-party targets.
Specifically, we present the following six attacks: 
\begin{itemize}
\item\textit{Listing active targets}
(\S\ref{sec:listing-active-targets}).
This is a privacy attack that can be used to track the list of active
running extensions and user's browsing
history, including URLs visited in an incognito window.
\item\textit{Running on regular tabs}
(\S\ref{sec:running-on-regular-tabs}).
This attack allows an extension to steal any user information contained inside most
browser tabs, including those in an incognito window, and evaluate
arbitrary code inside them to alter their behavior.
\item\textit{Running on security interstitial tabs}
(\S\ref{sec:running-on-security-interstitial-tabs}).
We show how extensions can abuse the Debugger API to
modify the contents of interstitial messages, including
critical security messages such as TLS error dialogs.
It also can be used to skip interstitials completely with no user interaction.
\item\textit{Running on WebUI tabs} (\S\ref{sec:running-on-webui-tabs}).
This attack extends the capabilities of the
second attack (\S\ref{sec:running-on-regular-tabs}). It allows
extensions to run on
internal tabs (\eg settings page), thus allowing the modification of critical
browser settings, including security ones.
\item\textit{Running on other extensions}
(\S\ref{sec:running-on-other-extensions}).
This attack allows extensions to debug any other running extension to
steal sensitive information
(such as plaintext passwords and credit card details stored in password
managers) or modify its normal
operation (\eg change the receiving wallet of a cryptocurrency transaction).
\item\textit{Attaching to the browser target}
(\S\ref{sec:attaching-to-the-browser-target}).
This attack interacts with a special high-privileged target from the CDP
that allows it to run on virtually any tab or extension and take full control
of the browser.
\end{itemize}

We confirm the feasibility of the
proposed attacks on every major Chromium-based browser,
including more privacy-focused solutions like Brave and Ungoogled Chromium.
All of our attacks share the same root cause, which we attribute to a mix of
questionable design decisions and excessive functionality granted to extensions
through CDP and the \texttt{debugger} permission. Overall, our attacks
exemplify the inherent tensions when reconciling a debugging tool, which by
definition has powerful capabilities, with an API exposed to
untrusted third-party code. This is a challenging design area and
we discuss the causes, impact, and potential mitigations in \S\ref{sec:discussion}.

\vspace{2mm}
\noindent\textbf{Disclosure.\xspace}
We disclosed all our findings to the Chromium Development Team along
with a practical Proof-of-Concept for each attack.
The attack from \S\ref{sec:listing-active-targets} and the vulnerable behavior
from \S\ref{sec:running-on-regular-tabs} were reported on
March 2022\footnote{Reported in \crbug{1301950}.}
and acknowledged to be a bug by Google. It was fixed in May 2022, when it landed in
Chromium Stable 102.
The attack from \S\ref{sec:running-on-security-interstitial-tabs} was reported
on November 2021\footnote{Reported in \crbug{1268445}.}
and was fixed in Chromium Stable 103,\footnote{See \crrev{3594083}.} when it was
assigned CVE-2022-2164.
The attack from \S\ref{sec:running-on-webui-tabs} was reported on
December 2021,\footnote{Reported in \crbug{1276500}.} marked as a duplicate
and merged with \S\ref{sec:running-on-other-extensions}.
As far as we know, this is the only bug that was previously known by the Chromium Team.
The attack from \S\ref{sec:running-on-other-extensions} was reported on
December 2021\footnote{Reported in \crbug{1276497}.} and
is still awaiting a fix from the Chromium Development Team.
The attack from \S\ref{sec:attaching-to-the-browser-target} was reported on
December 2021\footnote{Reported in \crbug{1276503}.}
and again for a different vulnerability on
March 2022.\footnote{Reported in \crbug{1301966}.}
The first report was marked as duplicated and merged with
\S\ref{sec:running-on-other-extensions}. The second report, while initially
flagged as a ``high-severity bug,'' was then classified as
``not-a-security-bug.'' We discuss this point in detail in
\S\ref{sec:discussion}.

\vspace{2mm}
\noindent\textbf{Research artifacts.\xspace}
Proof-of-Concepts for all our attacks are available
at \url{https://github.com/josemmo/chrowned}.
All of them use the Manifest V3 specification to account for the
approaching phase out of Manifest V2~\cite{google-mv2-sunset}.

\section{Background}
\label{sec:background}

Chromium is an open-source web browser project mostly
developed and maintained by Google.
Its codebase is the foundation for the Google Chrome browser,
and it is widely reused by many other popular browsers, including Microsoft
Edge, Brave, Samsung Internet or Opera.
This section provides technical background on Chromium extensions (\S\ref{sec:bg:extensions}), the
Chrome DevTools Protocol (\S\ref{sec:chrome-devtools-protocol}),
and Chromium's Debugger API (\S\ref{sec:cdp-debugger-api}).

\subsection{Chromium Extensions}
\label{sec:bg:extensions}

Chromium extensions are programs consisting of a set of files and assets
similar to those found in traditional web applications.
They are typically packaged into a custom file format known as ``CRX''
and then submitted to the Chrome Web Store---the official distribution platform
from Google---for distribution.
CRX packages are signed by extension developers using public-key cryptography
to provide integrity and authenticity.
As \S\ref{sec:threat-model} explains,
extensions can also be distributed outside the Chrome Web Store without
the need for signatures through \textit{sideloading} (\eg using ZIP archives).

Extensions must have a mandatory JSON file named \texttt{manifest.json}.
The manifest file contains basic metadata
about the extension (\eg name, version) and 
other properties, such as a background script (for running tasks or receiving events
independently of any opened tab) and content scripts (for injecting code onto
pages that match a given URL).

Chromium implements several security features to protect its users from
malicious extensions.
One of such features is isolation, which prevent websites and extensions alike
from running code on other execution contexts outside their scope.
Additionally, extensions must declare permissions in their manifest file to
access sensitive resources (like the browsing history) or break isolation and
run on tabs with a given URL (\ie through the use of host permissions).
Some permissions are considered more dangerous than others because of the
potential
harm that their abuse or misuse can cause to users. For this reason,
an extension containing one or more of these permissions will display a prompt
during install with brief warnings for each of
them~\cite{google-permission-warnings}.

Chromium's permission
model still presents important shortcomings, many of which have already been
fixed on platforms like Android~\cite{android-permissions}. For example,
once an extension is installed, it keeps unrestricted access to all
declared permissions and there is no simple way to revoke access
afterwards. Another limitation is transparency: the user warnings shown
in the install dialog communicate what a permission
enables at a technical level but it does not clearly convey how
impactful it can be for users' privacy and security if abused.

\subsection{Chrome DevTools Protocol}
\label{sec:chrome-devtools-protocol}

Chromium has a built-in component known as the Chrome DevTools Protocol
(CDP)~\cite{chrome-devtools-protocol} which allows web developers to
instrument and debug the browser to the fullest extent. Although this feature
can be abused as we demonstrate in this paper, it offers
useful features for software developers and researchers.
Selenium~\cite{selenium},
Puppeteer~\cite{puppeteer} and Playwright~\cite{playwright}
are three tools powered by CDP which are widely used
for End-to-End (E2E) testing of web applications or
automating web scraping.

Any application using the CDP communicates with the browser by sending
JSON-encoded messages to a WebSocket endpoint hosted by the Chromium process,
which is also used to receive events triggered by the browser.
For security and performance reasons, this endpoint
has to be enabled by adding the
\texttt{-{}-remote-debugging-port} command line flag when launching the
browser, followed by the port number the CDP server will be bound to
(port 9222 by convention)~\cite{chrome-devtools-protocol}.

\begin{figure}[t!]
  \centering
  \includegraphics[width=\columnwidth, frame]{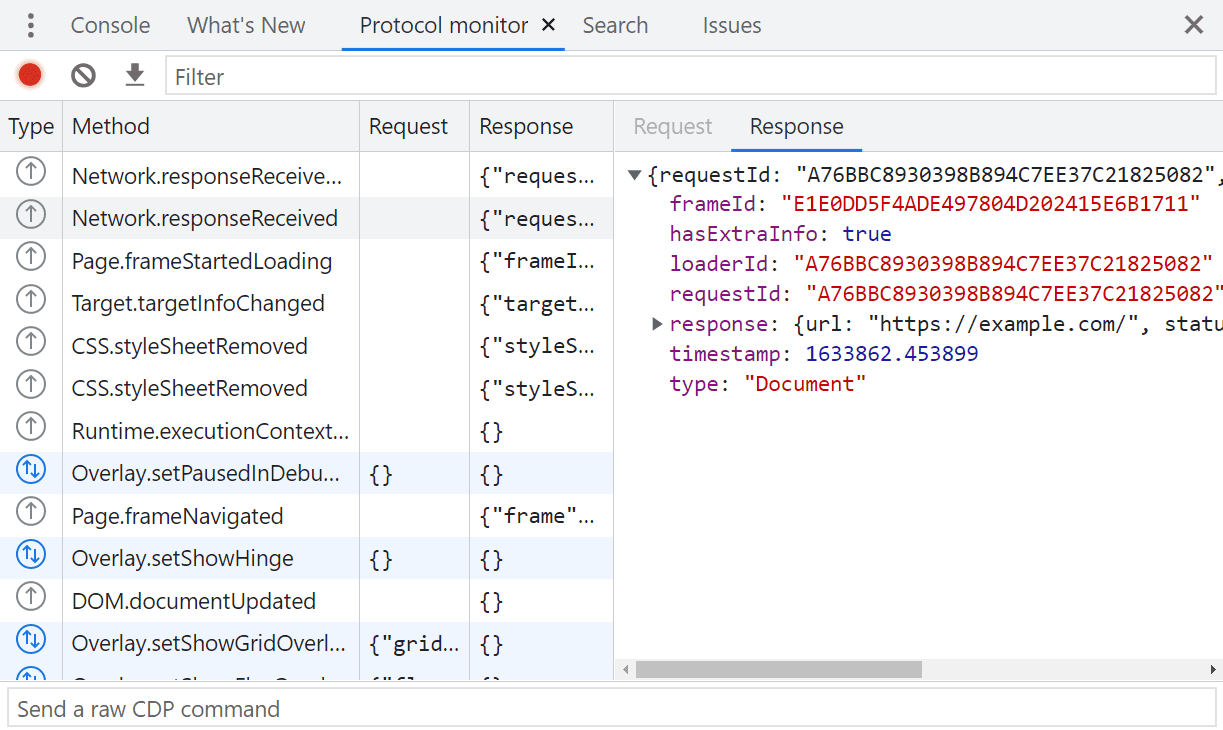}
	\caption{Protocol Monitor tab from DevTools UI window.
     }
  \label{fig:chromium-protocol-monitor}
\end{figure}

The Chrome DevTools UI~\cite{google-open-chrome-devtools} uses
the CDP under the hood to inspect and debug a target (\eg a tab).
Messages exchanged by the browser
(\textit{host}) and the DevTools UI (\textit{client}) can too be inspected
using the ``Protocol monitor'' drawer tab (see
Figure~\ref{fig:chromium-protocol-monitor}). This feature is
initially turned off but it can be manually
enabled from the ``Experiments'' section in
the DevTools UI settings menu.

To account for the isolation between tabs, background pages, service
workers,
and other worlds from the browser, CDP introduces the concept of
\textit{target}, which can be any of the former.
When a client wants to interact with a target, it first has to
\textit{attach} to it using the \texttt{Target.attachToTarget} command. It
will then receive a session ID to forward
CDP commands to the desired target.
This ensures that, for example, if we want to evaluate JavaScript code on
a particular tab, it will get executed only on our target and nowhere else.
By default, when opening a WebSocket to the CDP server, the client
attaches to what is known as the \textit{browser target}. This is a
special type of target that can list targets (\ie \texttt{Target.getTargets}),
attach to active ones and perform other instrumentation actions
affecting the entire browser.

The number of capabilities the protocol offers varies from version to version.
In a nutshell, the CDP can perform virtually any action a user will be able
to achieve with manual interaction (\eg clicking on a link or button).
Additionally, it gives finer control over browser internals that a regular user
cannot access, such as traffic interception or placing debugging breakpoints.
Specific capabilities provided by CDP include:
$(i)$ evaluate JavaScript code on the global object of a target (\eg \texttt{window});
$(ii)$ traverse, manipulate and add event listeners to the DOM tree of a target;
$(iii)$ intercept and modify network flows;
$(iv)$ list, edit, create and delete any entry from the browser cookie store;
$(v)$ grant or deny site permissions (\eg camera, geolocation) without requesting
user consent;
$(vi)$ take image screenshots of focused targets;
and $(vii)$ capture profiling snapshots of CPU, GPU
and memory, as well as metrics like code
coverage~\cite{google-analyze-runtime-performance}.

While most of these capabilities are truly sensitive, this is
expected from a development and debugging tool, and is not necessarily
concerning by itself. However, carelessly
exposing these capabilities to other parts of
the browser (\eg websites, extensions) might be a source of vulnerabilities.

\subsection{The Debugger extension API}
\label{sec:cdp-debugger-api}

Extensions cannot control the command line arguments the browser has been
launched with, which is a necessary step to activate the CDP WebSocket endpoint
(\S\ref{sec:chrome-devtools-protocol}).
Yet, extensions can still communicate with a version of the CDP with limited
capabilities through the use of the Debugger API~\cite{google-debugger}, which is
a substitute for virtually any other extension API:
by just requesting the \texttt{debugger} permission, extensions
can perform a wide range of sensitive actions
(\S\ref{sec:chrome-devtools-protocol}) without having to declare any
further permissions in the manifest.
In addition, it gives extensions access to exclusive functionality (\ie that no
other extension API offers).
Table~\ref{tab:features-table} provides a list of noteworthy Debugger API
features alongside their counterparts using other extension APIs, if any.

\begin{table}[t]
  \centering
  \caption{
    Summary of features provided through CDP commands and their equivalents in
    extension APIs.
  }
  \label{tab:features-table}
    \begin{tabular}{lll}
      \toprule
      \textbf{Feature}          & \textbf{CDP domain} & \textbf{Extension API}  \\
      \midrule
      Evaluate JS code          & \texttt{Runtime}    & \texttt{scripting}      \\
      Manipulate cookie store   & \texttt{Network}    & \texttt{cookies}        \\
      Intercept network traffic & \texttt{Network}    & \texttt{webRequest}     \\
      Take screenshots          & \texttt{Page}       & \texttt{desktopCapture} \\
      Modify site permissions   & \texttt{Browser}    & N/A                     \\
      Place breakpoints         & \texttt{Debugger}   & N/A                     \\
      Record execution trace    & \texttt{Tracing}    & N/A                     \\
      \bottomrule
    \end{tabular}
\end{table}

The Debugger API differs from the regular CDP in that it makes critical
protocol domains inaccessible. For example, the \texttt{Target} domain or even some
cherry-picked methods from other partially allowed domains.
The rationale behind this decision is presumably to prevent a rogue extension
from taking full control of the browser. We note that having a fully instrumented
environment is ideal when running E2E tests but it is a dangerous feature for
an end-user to have enabled, thus the need for a limited CDP agent for
extensions.
Given these limitations (mainly the absence of the \texttt{Target} domain),
an alternate API to list targets and attach to them is needed. Therefore, the
Debugger API provides JavaScript bindings to perform
these operations. In short, an extension has to follow three steps
to instrument a target:

\begin{enumerate}
  \item List all active targets with \texttt{chrome.debugger.getTargets()}
        to get the target, tab or extension ID of the one that will be
        instrumented.
  \item Call \texttt{chrome.debugger.attach()} to attach to the desired
        target. Immediately after attaching, the browser will start showing
        a notification to inform the user of this event.
  \item Use \texttt{chrome.debugger.sendCommand()} to send CDP commands to
        the debugged target and, optionally, add a listener to
        \texttt{chrome.debugger.onEvent} to be notified of CDP events.
\end{enumerate}

The Debugger API imposes limitations on what targets an extension can
attach to. While an extension can debug itself, it cannot instrument service
workers, background pages or tabs from other extensions, nor it can attach to
targets with a URL scheme other than ``http://'' or ``https://''.
We explore ways of bypassing these security mechanisms in
\S\ref{sec:attacks}.

\begin{figure}[t!]
  \centering
  \includegraphics[width=\columnwidth]{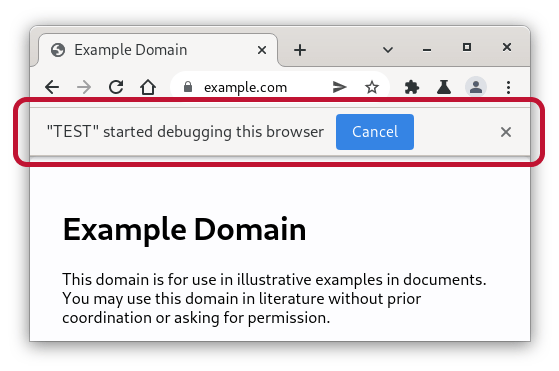}
	\caption{A sample debugger infobar in Chromium highlighted in \textcolor{purple}{red}.
  }
  \label{fig:chromium-infobar}
\end{figure}

\vspace{2mm}
\noindent\textbf{User awareness.\xspace}
\label{sec:cdp-debugger-inforbars}
When a debugging extension successfully attaches to a target, Chromium
notifies users by rendering an \textit{infobar} across all open
browser tabs~\cite{chromium-infobars} as shown in
Figure~\ref{fig:chromium-infobar}.\footnote{We note that debugger
infobars can be completely disabled if
Chromium is launched with the command line flag
\texttt{-{}-silent-debugger-extension-api}.}
This notification will not go away until there are no
attached debuggers left.
Optionally,
a user can click on the cancel button to force all debuggers from an extension
to be detached from their respective targets.
However, nothing prevents a malicious extension
from reattaching to the
previous targets immediately after the user cancels the infobar, and this
can happen without any user awareness (\S\ref{sec:running-on-regular-tabs}).

\section{Threat model}
\label{sec:threat-model}

Our fist threat model, named Threat Model A (TMA), assumes an attacker who can successfully
trick a user into installing a malicious extension that declares the
\texttt{debugger} permission, thus getting access to the capabilities described
in \S\ref{sec:chrome-devtools-protocol}.
This extension may be distributed through official platforms such as the Chrome
Web Store in the form of a signed CRX package~\cite{chromium-crx3-design-doc}.
Attacks \S\ref{sec:listing-active-targets} and \S\ref{sec:running-on-regular-tabs}
assume this threat model.

A second more restrictive threat model, named Threat Model B (TMB), requires users to
install extensions through sideloading.
Sideloading consists in loading an unpacked
extension~\cite{google-getting-started}, or drag-and-dropping a CRX
package or ZIP file over the extensions
management page (\ie \texttt{chrome://extensions}).
Because unpacked extensions and ZIP archives lack the cryptographic signature
that CRX files have, their contents are not integrity-protected nor their
authenticity can be verified. Therefore, they can spoof their ID
and impersonate other legitimate extensions. We introduce that feature in
\S\ref{sec:running-on-webui-tabs}).
However, as they are not CRX packages, they cannot be distributed
through the official Chrome Web Store.
Attacks \S\ref{sec:running-on-webui-tabs}, \S\ref{sec:running-on-other-extensions}
and \S\ref{sec:attaching-to-the-browser-target} use Threat Model B.

We note that sideloading is a feature intended for developers so it requires
users to enable ``Developer Mode'' on their browsers for it to work.
However, malicious actors can easily trick users into doing so or even silently
run a malicious PowerShell script that automates this process, a technique
that has already been observed in the
wild~\cite{therecord-north-korea, volexity-sharpext}.
A threat actor that gains execution capabilities outside the browser on a
victim's device might prefer sideloading an extension over dropping an executable
for ease of development and feature set completeness.
Browser extensions that work across different operating systems are easy to
develop, and the Debugger API lets attackers evaluate arbitrary JS code and
intercept network traffic in cleartext without the hassle of setting up a proxy
with a trusted self-signed certificate.

Although TMB is a more restrictive threat model, it is widely accepted as
realistic by the community. For
example, most numbered vulnerabilities (CVEs) involving a malicious
extension assume exactly this model (\eg \cite{cve-2018-6178, cve-2019-5768,
cve-2020-15973, cve-2021-21111, cve-2022-0107}), and it is also found in
previous academic works studying malicious browser
extensions~\cite{hulk, malicious-extensions-at-scale}.
Furthermore, sideloading is a common practice to install
extensions in regions where the Chrome Web Store is
not available and alternative marketplaces have
emerged~\cite{cws-alt-extmanager, cws-alt-crx4chrome, cws-alt-gugeapps}.
One notable example is China. There, sideloading is the only way
to install an extension because
the Chrome Web Store is not available despite Google Chrome
having the largest browser market share~\cite{statcounter-china}.

\section{Attacks}
\label{sec:attacks}

\begin{table*}[ht]
  \caption{Summary of attacks and their capabilities.
    All the attacks work in Chromium browser, Google Chrome, Microsoft Edge, Brave browser, Opera, Vivaldi and Ungoogled Chromium.}
  \label{tab:summary-table}

  \renewcommand{\arraystretch}{1.2}
  \aboverulesep=0ex
  \belowrulesep=0ex

  \resizebox{\textwidth}{!}{%
    \begin{tabular}{l|c|ccc|ccc|cccc}
      \toprule
      \multirow{2}{*}[-1em]{\textbf{Attack}}                                       & \multirow{2}{*}[-1em]{\makecell{\textbf{Theat}\\\textbf{Model}}} & \multicolumn{3}{c|}{\textbf{Opened tabs}}                                                     & \multicolumn{3}{c|}{\textbf{Running extensions}}                                  & \multicolumn{4}{c}{\textbf{Browser instance}}                                                                                                                               \\
                                                                                   &                                                            & \makecell{List} & \makecell{Evaluate\\ JS code*}             & \makecell{Intercept\\ traffic} & \makecell{List} & \makecell{Evaluate\\ JS code*} & \makecell{Intercept\\ traffic} & \makecell{Steal\\ browsing\\ cookies} & \makecell{Steal credit\\ cards and\\ passwords} & \makecell{Change\\ settings\\ and flags} & \makecell{Record\\ profiling\\ traces} \\
      \midrule
      Listing active targets (\S\ref{sec:listing-active-targets})                  & TMA                                                        & \CIRCLE         & \Circle                                    & \Circle                        & \CIRCLE         & \Circle                        & \Circle                        & \Circle                               & \Circle                                         & \Circle                                  & \Circle                                \\
      Running on regular tabs (\S\ref{sec:running-on-regular-tabs})                & TMA                                                        & \CIRCLE         & \LEFTcircle                                & \LEFTcircle                    & \CIRCLE         & \Circle                        & \Circle                        & \CIRCLE                               & \Circle                                         & \Circle                                  & \Circle                                \\
      Running on interstitials (\S\ref{sec:running-on-security-interstitial-tabs}) & TMB                                                        & \CIRCLE         & \LEFTcircle                                & \LEFTcircle                    & \CIRCLE         & \Circle                        & \Circle                        & \CIRCLE                               & \Circle                                         & \Circle                                  & \Circle                                \\
      Running on WebUI tabs (\S\ref{sec:running-on-webui-tabs})                    & TMB                                                        & \Circle         & \hspace{0.5pt} \CIRCLE \textsuperscript{†} & \Circle                        & \Circle         & \Circle                        & \Circle                        & \Circle                               & \CIRCLE                                         & \CIRCLE                                  & \Circle                                \\
      Running on extensions (\S\ref{sec:running-on-other-extensions})              & TMB                                                        & \CIRCLE         & \LEFTcircle                                & \LEFTcircle                    & \CIRCLE         & \CIRCLE                        & \CIRCLE                        & \CIRCLE                               & \Circle                                         & \Circle                                  & \Circle                                \\
      Attaching to browser (\S\ref{sec:attaching-to-the-browser-target})           & TMB                                                        & \CIRCLE         & \CIRCLE                                    & \CIRCLE                        & \CIRCLE         & \CIRCLE                        & \CIRCLE                        & \CIRCLE                               & \CIRCLE                                         & \CIRCLE                                  & \CIRCLE                                \\
      \bottomrule
    \end{tabular}%
  }

  \vspace{-5pt}
  \begin{small}
    \begin{flushleft}
      \begin{tabular}{p{0.41\textwidth}p{0.59\textwidth}}
        \makecell[tl]{
          \Circle~~Not capable at all.\\
          \LEFTcircle~~Only on regular tabs and interstitials.\\
          \CIRCLE~~Fully capable of.
        } &
        \makecell[tl]{
          {*}~~Useful for stealing data from a target and manipulating its behavior.\\
          {†}~~Except tabs from incognito windows.
        }
      \end{tabular}
    \end{flushleft}
  \end{small}

\end{table*}

We next introduce six attacks that exploit design vulnerabilities
in the Debugger API and in the logic for granting
restricted capabilities to highly privileged extensions.
Attacks are grouped by threat model (first TMA, then TMB) and sorted
by severity in ascending order.

\vspace{2mm}
\noindent\textbf{Methodology.}
We performed a systematic manual code review process to find issues in the
design and implementation of the Debugger API.
To assess its security, we defined a set of security requirements that
an hypothetically
secure extension API must comply with based on the official CRX API Security
Checklist~\cite{chromium-security-checklist}.
This is a public document used by Chromium developers as a baseline to follow
best security practices~\cite{chromium-security-education}.

To do so, we systematically review the JavaScript bindings exposed by the
\texttt{chrome.debugger} object~\cite{code-debugger-api}. For each binding,
we flag its source code and that of the browser components it depends on.

Then, we thoroughly inspect the tainted source code to find violations of
the following requirements:

\begin{itemize}
  \item \textit{SR01 -- User Awareness.}
  Users must be clearly informed of the implications of an extension making
  use of the Debugger API (\eg during installation) and notified when an
  extension is actively using it (\eg using infobars).

  \item \textit{SR02 -- Isolation.}
  The Debugger API must respect browser profiles and honor users' privacy choices
  (\eg incognito mode) and
  restrict the scope of extensions using it accordingly.

  \item \textit{SR03 -- Access Control.}
  The Debugger API must enforce rules to prevent extensions from accessing
  sensitive
  resources (\eg browsing history) or targets (\eg tabs) outside their
  supposed reach.

  \item \textit{SR04 -- Spoofing Avoidance.}
  Extensions should not be able to modify critical parts of the browser UI
  (\eg settings page) to stop them from misleading users into performing a
  given action.
  
\end{itemize}

To identify such violations, we looked for security checks or preconditions
that the Debugger API should comply with.
For instance, to satisfy SR01, a call to display an infobar with a localized
message is likely expected when an extension attaches to a target.
Once a requirement violation is found, we manually verified its potential
exploitability by implementing a prototype extension and testing its
effectiveness against the latest Chromium Stable release at the time.

Table~\ref{tab:summary-table} summarizes the attacks that we found
using this methodology and their impact, and
Table~\ref{tab:requirements-table} shows violations of the above Security
Requirements for each attack.
The order in which we introduce the attacks goes from the least to the most
impactful, starting from merely listing the opened tabs and running extensions
(\S\ref{sec:listing-active-targets}), then gaining access to evaluate arbitrary
code on each of those targets (\S\ref{sec:running-on-regular-tabs},
\S\ref{sec:running-on-security-interstitial-tabs},
\S\ref{sec:running-on-webui-tabs}, \S\ref{sec:running-on-other-extensions}),
and finally being able to take control of the browser
(\S\ref{sec:attaching-to-the-browser-target}).
Note that some of our attacks achieve similar results. We include them nonetheless as they
are independent from one another (\ie they are based on different flaws, thus
mitigating one attack will not affect the other).

\vspace{2mm}
\noindent\textbf{Prevalence.\xspace}
We are able to reproduce our attacks in all major
Chromium-based browsers (Google Chrome, Microsoft Edge, Opera and Vivaldi),
including those with a special focus on privacy
(Brave browser and Ungoogled Chromium~\cite{ungoogled-chromium}).
We note that we have not tested these attacks on mobile versions of Chromium
(\eg Google Chrome for Android) as 
there is no
straightforward way of installing extensions in those builds as of
this writing.
We provide Proofs-of-Concept (PoCs) for all
of the attacks in the artifacts associated with this paper for independent
validation. These PoCs are
fully-commented extensions based on our threat model.
They showcase
various real-world scenarios that make use of our findings.

\begin{table}[t]
  \centering
  \caption{Security Requirement violations per attack.}
  \label{tab:requirements-table}

  \newcommand{\cmark}{\xspace}
  \newcommand{\xmark}{\ding{55}}
  \newcommand{\xxmark}{\textcolor{gray}{\hspace{0.5pt} \xmark \textsuperscript{‡}}}

    \begin{tabular}{lccccc}
      \toprule
      \multirow{2}{*}{\textbf{Attack}}                  & \multicolumn{2}{c}{\textbf{SR01}}             & \multirow{2}{*}{\textbf{SR02}} & \multirow{2}{*}{\textbf{SR03}} & \multirow{2}{*}{\textbf{SR04}} \\
                                                        & \makecell{Install-time} & \makecell{Run-time} &                                &                                &                                \\
      \midrule
      \S\ref{sec:listing-active-targets}                & \xmark                  & \xmark              & \xmark                         & \xmark                         & \cmark                         \\
      \S\ref{sec:running-on-regular-tabs}               & \xmark                  & \xxmark             & \xmark                         & \xmark                         & \cmark                         \\
      \S\ref{sec:running-on-security-interstitial-tabs} & \xmark                  & \xxmark             & \xmark                         & \xmark                         & \xmark                         \\
      \S\ref{sec:running-on-webui-tabs}                 & \xmark                  & \xxmark             & \cmark                         & \xmark                         & \xmark                         \\
      \S\ref{sec:running-on-other-extensions}           & \xmark                  & \xxmark             & \xmark                         & \xmark                         & \cmark                         \\
      \S\ref{sec:attaching-to-the-browser-target}       & \xmark                  & \xxmark             & \xmark                         & \xmark                         & \xmark                         \\
      \bottomrule
    \end{tabular}

  \vspace{-2pt}
  \begin{small}
    \begin{flushleft}
      \hspace{10pt} \xmark~~Denotes a violation of a security requirement.\\
      \hspace{10pt} \textcolor{gray}{\xmark\textsuperscript{‡}}~Only a violation if infobars have been disabled.
    \end{flushleft}
  \end{small}

\end{table}

\subsection{Listing active targets}
\label{sec:listing-active-targets}

The Debugger API is implemented in such a way
 that extensions have to call the
\texttt{chrome.debugger.attach()} function to start instrumenting a
particular target using the CDP (see \S\ref{sec:cdp-debugger-api}).
To indicate the target, its target ID must be
passed as a parameter to this function.
Yet, extensions can extract a list of running target IDs by calling
\texttt{chrome.debugger.getTargets()}.

\vspace{2mm}
\noindent\textbf{Attack vector.}
The \texttt{chrome.debugger.getTargets()} function lists not only the
targets an extension is allowed to attach to (usually tabs with a URL
beginning with ``http://'' or ``https://''), but also other targets the
extension is not supposed to be capable of debugging.
Any extension declaring the \texttt{debugger} permission in its manifest can
call this function from both content scripts and background pages without
requiring any user interaction. This could result in a privacy abuse
as it can expose sensitive data such as users' browsing history, thus violating SR03.
Being able to access this type of sensitive
information raises privacy concerns as it could be linked to
user identities (\eg by also harvesting cookies or other Personal
Identifiable Information).

\vspace{2mm}
\noindent\textbf{Impact.}
Alongside the target ID, the Debugger API also grants access to the URL of the
target, the page title and even the favicon URL.
A malicious extension can call \texttt{chrome.debugger.getTargets()} to
easily and accurately monitor the set of opened tabs and the list
of websites visited by
the user, including those from incognito windows. Note that the
malicious extension does not need to be granted permission to run in the
former context, thus breaking SR02.
Additionally, because service workers and background pages are also listed,
an attacker can also get the list of running extensions.\footnote{
  Similarly, the Management API~\cite{google-management} provides a method for
  listing installed extensions, though it requires declaring an
  additional permission.
}
Access to installed extensions can be used for improving
device fingerprinting
techniques, as in some cases is enough
to unequivocally identify a user~\cite{extension-resources-control-policies,
xhound, fingerprinting-in-style}.
Since there is no rate-limit to getting the list of targets, an attacker
could keep polling this information every few seconds or less to detect
changes in enabled extensions or page navigation events, as proposed in
Listing~\ref{lst:attack-1-example}.
An important
remark is that SR01 is also violated as no debugger infobar
(see \S\ref{sec:cdp-debugger-inforbars}) is shown when calling the
\texttt{chrome.debugger.getTargets()} method, thus making this attack fully
silent.
We provide a PoC within the artifacts to detect running extensions
that solely uses the Debugger API.

\begin{lstlisting}[
  float=t!,
  caption={JavaScript code for polling running targets every second.},
  label={lst:attack-1-example}
]
const visited = new Set();
setInterval(async () => {
  const targets = await chrome.debugger.getTargets();
  for (const target of targets) {
    if (visited.has(target.url)) {
      // Ignore previously visited URLs
      continue;
    }
    visited.add(target.url);
    console.log({time: Date.now(), url: target.url});
  }
}, 1000);
\end{lstlisting}

\subsection{Running on regular tabs}
\label{sec:running-on-regular-tabs}

\textit{Regular tabs} are the least privileged targets a CDP client can
attach to. They use either the ``http://'' or ``https://'' scheme.
We note that the ``ftp://'' scheme is unsupported since October
2021~\cite{google-cr95-deprecations}, and ``file://'' is restricted by
default and requires the user to manually opt in~\cite{google-match-patterns}.

\vspace{2mm}
\noindent\textbf{Attack vector.}
Instrumenting a regular tab is trivial. The extension only needs to declare
the \texttt{debugger} permission in its manifest and then attach to the
desired tab using its tab ID or target ID. Both IDs can be known
with the methods described in \S\ref{sec:listing-active-targets}.
Once attached, CDP messages can
be exchanged (see \S\ref{sec:cdp-debugger-api}) as
long as the extension does not get detached from the target, which can happen
programmatically by purposely calling \texttt{chrome.debugger.detach()},
or when the tab is closed or changes the URL to a restricted page (\ie
outside the scope of regular tabs).
Because there is no finer control over this API, an extension with the
\texttt{debugger} permission will have unrestricted access to the Debugger API
in its entirety. 
This breaks SR03 as it can be abused to access sensitive resources that otherwise would
require additional permissions (\eg the browser cookie store).

After attaching to a target, an infobar is shown to the user giving the
option to force-detach the debugging extension (\S\ref{sec:cdp-debugger-inforbars}).
However, during our research we find that extensions can reattach immediately
afterwards and that no rate-limit protection exists against this behavior.
In practice, to the user this would look like the cancel button of an
infobar does nothing, rendering it ineffective due to the
immediacy at which extensions can re-attach.
In addition, due to an incomplete security check, an extension can also attach to
regular tabs from incognito windows,
even when it has not been granted permission to do so, which is the default
behavior.
While trying to attach to an incognito tab by providing the tab ID will result in
a \textit{``No tab with given id''} error, no
additional input verification is performed when we use its target ID,
thus letting the extension debug tabs from incognito windows and violating SR02.

\vspace{2mm}
\noindent\textbf{Impact.}
As mentioned in \S\ref{sec:cdp-debugger-api}, the Debugger API is a
general substitute for virtually any other extension API.
In an nutshell, with this API, a malicious extension can:

\begin{itemize}
  \item\textit{Steal sensitive user information.}
By being able to manipulate the DOM or evaluate arbitrary JavaScript expressions,
the extension can read email addresses, passwords, credit card numbers and other
Personal Identifiable Information (PII) already accessible to the regular tab.
It is also possible to monitor all network traffic sent by the regular tab to
which the extension is
attached to for the same purpose.
Additionally, the \texttt{Network.getAllCookies} command lets extensions exfiltrate any
cookie stored in the browser regardless of the URL of the debugged regular tab,
thus providing a historical perspective of users' browsing habits
and even leaking sensitive data encoded in the cookies as
Listing~\ref{lst:attack-2-example} shows.

  \item\textit{Manipulate runtime behavior.}
Apart from stealing data, evaluating arbitrary code on regular tabs is also
useful for modifying the UI of a web application or changing its intended
behavior.
For instance, an attacker may inject a JavaScript file on a banking app to
alter the recipient of a wire transfer while hiding this information from the
user.

  \item\textit{Modify site permissions.}
The \texttt{Browser.grantPermissions} command can grant any site access to
privacy-sensitive resources (\eg microphone) without further user consent,
bypassing the need for additional permission in an extension's manifest.

\end{itemize}

We verify the previous capabilities by creating several extensions that
use the Debugger API to achieve different goals. As a PoC, we provide an
extension that attaches to tabs from incognito windows, and another one
that list the metadata for all cookies stored in the browser.

\begin{lstlisting}[
  float=t!,
  caption={JavaScript code for reading all browser cookies.},
  label={lst:attack-2-example}
]
// Attach to ourselves (any regular tab will do)
const targets = await chrome.debugger.getTargets();
const targetId = targets.find(target => {
  return target.url === document.location.href;
}).id;
await chrome.debugger.attach({targetId}, "1.3");

// Get all cookies
const res = await chrome.debugger.sendCommand(
  {targetId},
  "Network.getAllCookies"
);
console.log(res.cookies);
\end{lstlisting}

\subsection{Running on security interstitial tabs}
\label{sec:running-on-security-interstitial-tabs}

\begin{figure}[t!]
  \centering
  \includegraphics[width=0.7\columnwidth, frame]{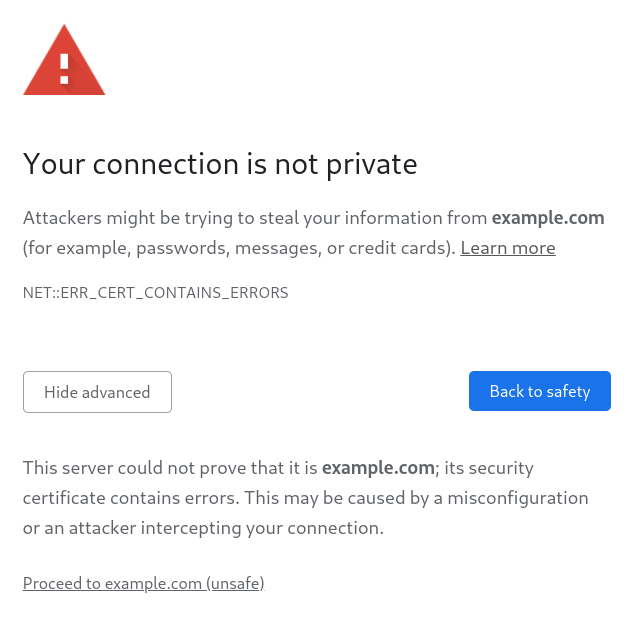}
  \caption{A privacy error interstitial in Chromium.}
  \label{fig:chromium-interstitial}
\end{figure}

Chromium comes with a built-in security feature known as \textit{Safe Browsing}
for blocking known phishing and malware sites~\cite{chromium-safe-browsing}.
When users enable this feature, the browser will block navigation to URLs
included in a blocklist and instead display an \textit{interstitial}
dialog asking the user to confirm a dangerous action (\eg continue anyway and visit
the site) or abort and go back to the previous page.
Interstitials are also used for showing SSL/TLS warnings like
an expired or invalid certificate as shown in Figure~\ref{fig:chromium-interstitial},
or when connecting to a public network behind a
Captive Portal, among other use cases.

\vspace{2mm}
\noindent\textbf{Attack vector.}
Given the sensitive nature of interstitials, these ``loud'' dialogs are always
served by Chromium at \texttt{chrome-error://chromewebdata/}, which is a
special unreachable URL (\ie it cannot be visited by typing it on the
browser's address bar).

Serving such interstitials like so presents a security advantage:
because the target is no longer a regular tab (\ie it
has a different scheme),
extensions cannot theoretically run on it, thus preventing a malicious actor from, for
instance, using the \texttt{scripting} extension API to evaluate arbitrary
JavaScript code on the page.
However, this restriction does not apply to the Debugger API, which can
attach to targets with the former unreachable URL. This lack of access
control to such a sensitive class of resources enables several attacks that
can impersonate or modify a security interstitial.

\vspace{2mm}
\noindent\textbf{Impact.}
Given that extensions can use the Debugger API to attach to a security
interstitial tab, an attacker can use this capability to:

\begin{itemize}
  \item\textit{Impersonate interstitials.}
Extensions can evaluate arbitrary code to modify the appearance of these
notifications to show a different message or content, therefore modifying
the browser UI and violating SR04.
Performing this attack is as trivial
as querying the DOM nodes we want to modify (\eg using
\texttt{document.querySelector}) and then changing their HTML or text contents.
A clear malicious use case for this is a phishing attack, given that it can
produce an interstitial
with the same look and feel of the ones triggered by Chromium to trick
the user into performing some dangerous action, like downloading an executable
as Figure~\ref{fig:chromium-phishing} shows.

\begin{figure}[t!]
  \centering
  \includegraphics[width=\columnwidth]{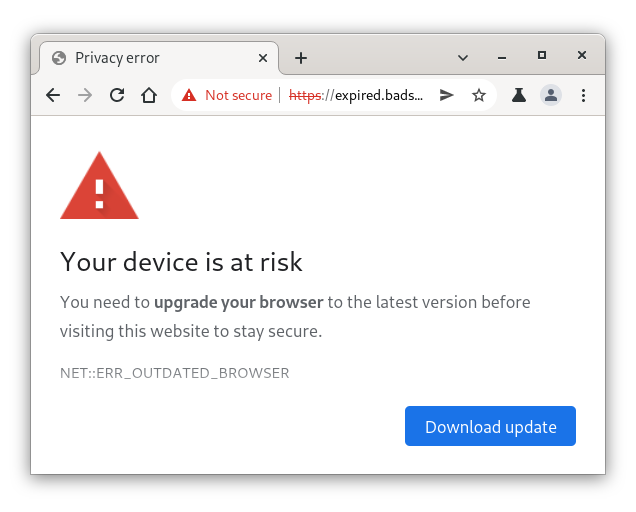}
  \caption{An impersonated security interstitial used to trick the user
     into downloading a malicious executable under the pretext that a
     browser update is needed.}
  \label{fig:chromium-phishing}
\end{figure}

  \item\textit{Skip interstitials.}
An attacker can also force the browser to automatically skip interstitials
on some websites by making the extension
programmatically click on the ``Proceed'' button or calling
\texttt{certificateErrorPageController.proceed()}.
This could be used by a malicious extension to skip TLS warnings or error
messages (\eg related to certificate validation), which can facilitate
TLS man-in-the-middle (MITM) attacks using forged certificates. The same
strategy can be applied to sideload malware by skipping Safe Browsing.

\end{itemize}

To demonstrate this attack, we created a PoC
extension that automatically bypasses
security interstitials whenever it encounters one.

\subsection{Running on WebUI tabs}
\label{sec:running-on-webui-tabs}

A vast portion of the User Interface (UI) in Chromium is implemented using web
pages. A well-known case are, for example, the browser settings, which can be
accessed by navigating to the \texttt{chrome://settings} URL.
There are well over 70
different URLs using the internal ``chrome://'' scheme,
ranging from the bookmarks menu to the
browsing history. Google calls
these \textit{Chrome UI} or \textit{WebUI} pages~\cite{chromium-webui-explainer}.
An almost complete list of WebUI URLs can be found on
\texttt{chrome://about}.

These pages run in a higher-privileged context with more capabilities than
regular tabs, which are provided through
$(i)$ internal extension APIs that are granted based on the URL; and
$(ii)$ message passing communication using Mojo JavaScript
bindings~\cite{chromium-mojo-docs}.
For example, the \texttt{chrome://extensions} page, responsible for letting
the user manage the installed browser extensions, has access to an
internal extension API (\ie \texttt{developerPrivate}) for that very purpose.

For obvious reasons, extensions cannot run on WebUI pages as that would
have devastating consequences for the browser's security model.
However, on March 2013 the experimental flag \texttt{extensions-on-chrome-urls}
was added to Chromium to circumvent this limitation and allow extensions that
explicitly declared the \texttt{chrome://*} permission to access WebUI
pages~\cite{chromium-cr-chrome-urls}.
Later on, in 2018, a security report flagging this issue
triggered a response from the
Chromium Security Team~\cite{gerrit-webui-harcoded} that limited the
capabilities of the flag by blocking extensions using the Debugger API from
attaching to WebUI targets regardless of the value of the previous flag
as
Listing~\ref{lst:mayAttachToURL} shows.

\begin{lstlisting}[
  float=t!,
  caption={%
    Code to prevent extensions from attaching to WebUI pages
    (adapted from~\cite{code-debugger-api}).%
  },
  label={lst:mayAttachToURL}
]
bool ExtensionDevToolsClientHost::MayAttachToURL(
  const GURL& url,
  bool is_webui
) {
  if (is_webui) {
    return false;
  }
  // [...]
  return true;
}
\end{lstlisting}

\vspace{2mm}
\noindent\textbf{Attack vector.}
Not being able to use the Debugger API does not imply that extensions
cannot run on WebUI tabs. An extension can still use the \texttt{tabs} or
\texttt{scripting} APIs to evaluate JavaScript expressions when the
previous flag is enabled.
In fact, the official ``Screen Reader'' extension~\cite{cws-screen-reader} has
access to the ``chrome://'' scheme without needing it to be enabled at all.
This extension is an accessibility tool targeted towards users
with visual impairments that can read aloud the contents, buttons, menus and
other elements of a page.
We find other instances of security concessions made in favor of usability in
platforms such as Android, where it has proven to be abusable by
attackers~\cite{android-ui-security-undermined, kindness-is-a-risky-business}.
Because a significant part of the browser UI is implemented using web pages,
the extension has to be able to interact with WebUI tabs too to read its
contents. To accomplish it, the Chromium developers made an
exception for this particular extension
and granted it privileges to run on the former higher-privileged targets
without requiring any further flags.

This backdoor access is implemented by defining an allowlist
solely containing the Screen Reader extension ID, which is queried
every time an extension wants to run on a given URL to determine if it is
restricted
or not (see Listing~\ref{lst:CanExecuteScriptEverywhere}).

\begin{lstlisting}[
  float=t!,
  caption={%
    Code to restrict the URLs an extension can run on
    (adapted from~\cite{code-permissions-data}).%
  },
  label={lst:CanExecuteScriptEverywhere}
]
bool PermissionsData::CanExecuteScriptEverywhere(
  const ExtensionId& extension_id,
  ManifestLocation location
) {
  if (location == ManifestLocation::kComponent) {
    return true;
  }
  const auto& allowlist = GetScriptingAllowlist();
  return base::Contains(allowlist, extension_id);
}

bool PermissionsData::IsRestrictedUrl(
  const GURL& document_url,
  std::string* error
) const {
  // Grant access to Screen Reader extension
  if (CanExecuteScriptEverywhere(ext_id_, loc_)) {
    return false;
  }

  // Block access to WebUI without flag
  bool allow_on_chrome_urls = ForCurrentProcess()
    ->HasSwitch(kExtensionsOnChromeURLs);
  if (
    document_url.SchemeIs(kChromeUIScheme) &&
    !allow_on_chrome_urls
  ) {
    return true;
  }

  // Block access to other extensions without flag
  if (
    document_url.SchemeIs(kExtensionScheme) &&
    document_url.host() != ext_id_ &&
    !allow_on_chrome_urls
  ) {
    return true;
  }

  // [...]
  return true;
}
\end{lstlisting}

\begin{figure}[t!]
  \centering
  \includegraphics[width=\columnwidth]{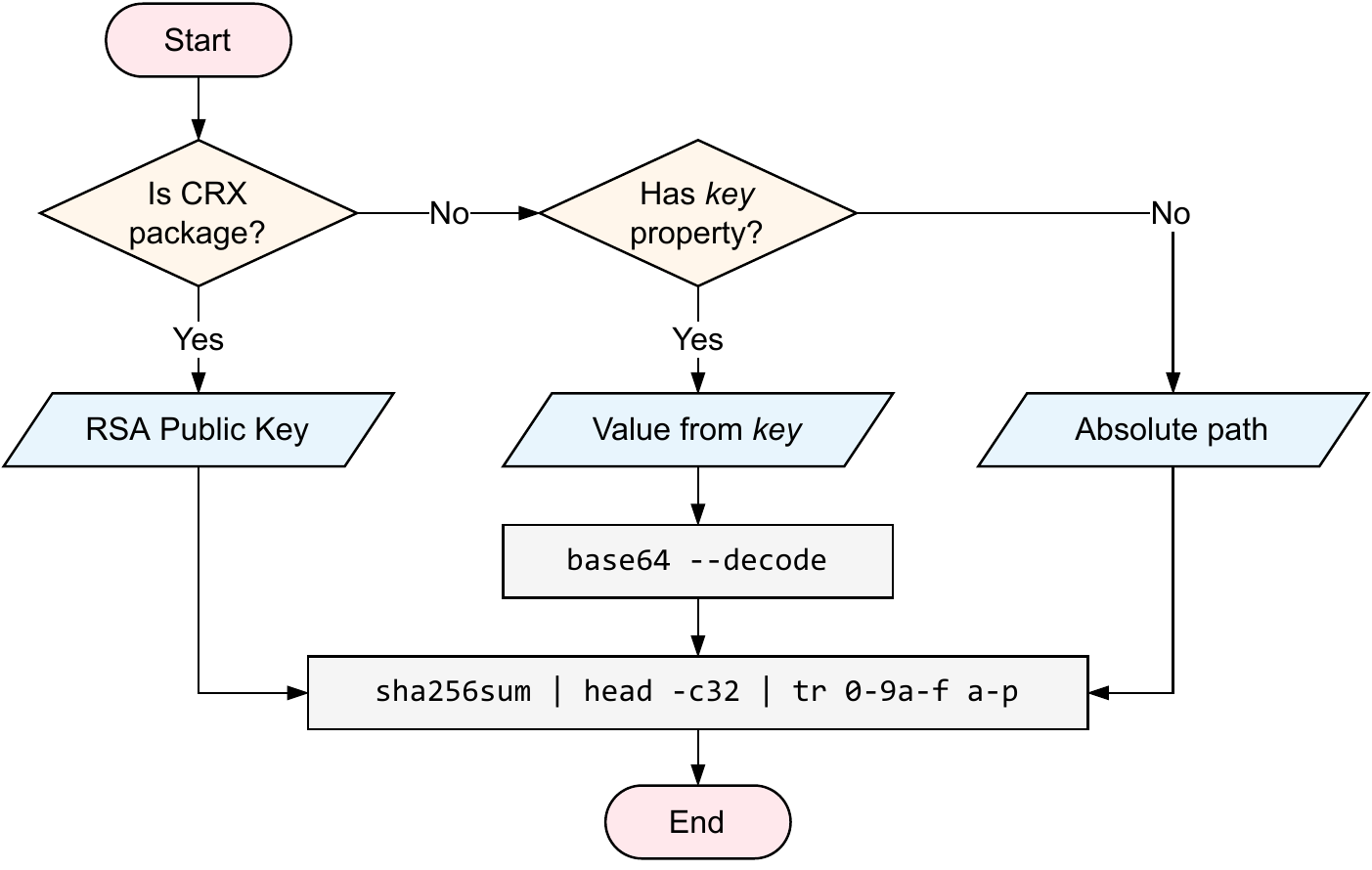}
  \caption{Flowchart for calculating extension IDs.}
  \label{fig:extension-id-flowchart}
\end{figure}

Using a hardcoded allowlist for this purpose is not a robust nor secure
design choice as the ID of an extension can be impersonated.
A \textit{clone} extension with the same ID as the Screen Reader
extension can take advantage of its privileged position within the browser
and acquire its additional capabilities.
Unfortunately, this is rather trivial to perform.
Figure~\ref{fig:extension-id-flowchart} shows how
Chromium calculates extension IDs by taking a
DER-encoded RSA public key and producing a SHA-256 digest of its bytes, with
the added distinction that the output hash in hexadecimal form uses an ``a'' to
``p'' alphabet instead of the standard ``0-9'' and ``a-f''.
This digest is then cropped to
return only the first 32 characters~\cite{robwu-calculate-extension-id}.
To get this public key for the ID generation, the browser looks at the header
of the extension's signed CRX package it came in and additionally performs
an integrity check against its signature~\cite{chromium-crx3-design-doc}.
However, if the extension is sideloaded from a local directory (unpacked
extension) or a ZIP archive instead of using a CRX package (\eg downloaded from the
Chrome Web Store), this public key information is not available. In
those cases, the ID is derived from the absolute path the extension was loaded
from unless there is a \texttt{key} property in the extension's
\texttt{manifest.json} file (see Listing~\ref{lst:attack-4-example}).
If the property exists, whatever value it has will be used as the extension's
Base64-encoded public key~\cite{google-manifest-key}, thus making it possible to
produce clone extensions.

\begin{lstlisting}[
  float=t!,
  caption={Sample manifest impersonating the Screen Reader extension.},
  label={lst:attack-4-example}
]
{
  "manifest_version": 3,
  "name": "Clone extension",
  "version": "0.0.1",
  "key": "MIGfMA0GCSqGSIb3DQEBAQUAA4GNADCBiQKBgQDEGB
    i/oD7Yl/Y16w3+gee/95/EUpRZ2U6c+8orV5ei+3CRsBsoXI
    /DPGBauZ3rWQ47aQnfoG00sXigFdJA2NhNK9OgmRA2evnsRR
    bjYm2BG1twpaLsgQPPus3PyczbDCvhFu8k24wzFyEtxLrfxA
    GBseBPb9QrCz7B4k2QgxD/CwIDAQAB"
}
\end{lstlisting}

Clone extensions have some limitations. Without the private key of
the impersonated extension, we cannot produce a valid signed CRX package and
upload it to the Chrome Web Store.
Therefore, the easiest alternate distribution method is to create a ZIP archive
of the clone extension and convince the user to enable Developer Mode and then
drag-and-drop the ZIP file over the \texttt{chrome://extensions} page to
install it.
This is one way of \textit{sideloading} extensions into Chromium and,
as discussed in \S\ref{sec:threat-model}, it is a usual
threat model for malicious extensions.

\vspace{2mm}
\noindent\textbf{Impact.}
Allowing extensions to run arbitrary code on WebUI pages violates both SR03 and
SR04 because these pages have access to sensitive resources (\eg browsing
history) and control parts of the browser UI.
Some applications of this attack include:
\begin{itemize}
  \item\textit{Modify browser settings.}
A clone extension can evaluate
JavaScript on the \texttt{chrome://settings} page to read or tamper
with the browser settings. This can be done without user interaction.

  \item\textit{Steal passwords and credit cards.}
Chromium comes with a built-in password and payment methods manager.
Because plain text passwords and credit card details are accessible through
the settings page, an attacker can obtain this information in plain text.\footnote{
  In some Operating Systems like Windows, Chromium stores passwords encrypted
  with the logged user credentials. For this reason, leaking passwords
  might require users to authenticate themselves to proceed.
}

  \item\textit{Read GAIA ID.}
Every Google account is identified by a unique
\textit{Google Accounts and ID Administration ID}, or GAIA ID for
short~\cite{google-gaia-id}.
This identifier is stored by the browser when a Google account has logged in
and can be leaked from the \texttt{chrome://signin-internals/} page.

  \item\textit{Modify browser flags.}
Some browser settings (like the flag mentioned in this section) are not
intended to be changed by regular users because of security reasons.
However, since these are listed on \texttt{chrome://flags}, they can be
arbitarily changed by a malicious extension.

  \item\textit{List omnibox predictors.}
To provide relevant autocomplete suggestions, Chromium keeps track locally of
all text typed into the search bar (\ie \textit{omnibox}~\cite{chromium-omnibox}).
The page \texttt{chrome://predictors/} exposes this data as a WebUI tab, making
it accessible to an attacker.

\end{itemize}

Our PoC demonstrates how this attack allows
obtaining the details of payment methods found
in \texttt{chrome://settings/payments},
one of the browser's settings pages.

\subsection{Running on other extensions}
\label{sec:running-on-other-extensions}

Similar to WebUI pages, which are served from the privileged ``chrome://''
scheme, extensions also run in a restricted environment, each having their own
dedicated base URL in the form of \texttt{chrome-extension://<extension-id>}.
In practice, this implies that an extension can access any assets and run on targets
belonging to its base URL (like its background page or service worker), but
it is not allowed to do the same with other extensions.
Despite running in isolation from each other, extensions can communicate with
other extensions, tabs and background scripts through message
passing~\cite{google-message-passing}. While message communication can be used
to evaluate code on a vulnerable target~\cite{doublex}, extensions can never
directly run on a third-party extension using browser extension APIs like they
can with regular tabs.

\vspace{2mm}
\noindent\textbf{Attack vector.}
The same experimental flag used in \S\ref{sec:running-on-webui-tabs}
to run on WebUI tabs grants access to the ``chrome-extension://'' scheme
(Listing~\ref{lst:CanExecuteScriptEverywhere}).
However, there is no way to declare a host permission with this scheme in the
manifest file of an extension as that will throw a validation warning and
the browser will ignore the permission altogether.
A crucial aspect here is that, while most browser extension APIs, like
\texttt{tabs} and \texttt{scripting}, will verify host permissions before
running on a target, the Debugger API implicitly grants access to any URL
an extension is authorized to access. For this reason, it only checks
whether a URL is restricted or not by calling the
\texttt{PermissionsData::IsRestrictedUrl()} method.
Because this method effectively considers a URL unrestricted when the
\texttt{extensions-on-chrome-urls} flag is enabled, the Debugger API can
be used to attach to any target from any extension.
This can be used as an alternative exploitation method without requiring
sideloading nor impersonating the Screen Reader extension using the technique
described in \S\ref{sec:running-on-webui-tabs}.

\vspace{2mm}
\noindent\textbf{Impact.}
A malicious extension exploiting this technique can attach to an existing
legitimate extension to:
\begin{itemize}
  \item\textit{Steal sensitive information.}
By attaching to another extension, an
attacker can evaluate JavaScript code to steal passwords, PGP private keys or
other sensitive information, hence breaking SR03.
A clear target for this are password manager extensions.

  \item\textit{Manipulate the extension behavior.}
Similar to \S\ref{sec:running-on-regular-tabs}, debugging an extension
can be used to change its appearance and behavior. This allows, for example,
stealing funds by modifying the receiving address of a cryptocurrency transaction
initiated with MetaMask~\cite{metamask}.

\end{itemize}

This attack presents an interesting property that eases its
distribution and amplifies its impact.
In the past, there have been instances of legitimate extensions going
rogue. One common cause is developers selling their published items to
other companies
that then push a malicious update to steal private user information. These
extensions operate for a limited time on a trusted marketplace (\eg the
Chrome Web Store) before they get spotted and
delisted from the market~\cite{xda-great-suspender-hijacked, threatpost-copyfish-hijacked}.
Another popular approach to distribute malware is to publish a fake extension
impersonating or cloning a legitimate one to trick a user into trusting the
former~\cite{ghacks-microsoft-authenticator-fake}. This is accomplished either
by having a very similar UI
or just by building a modified version of the original extension with some
patches applied to add the malicious functionality.
The former approach has been seen in
a recent operation attributed to North Korea that made the news in January
2022~\cite{therecord-north-korea}. 
The attack described in this section would not require repackaging fake
extensions nor buying existing ones. Instead, an attacker just needs to
distribute a malicious extension that will attach to as many legitimate
targets (\ie extensions) as needed to monitor and manipulate their behavior.

To demonstrate the stealing capabilities, we created a PoC extension that
attaches to LastPass~\cite{cws-lastpass} and
Mailvelope~\cite{cws-mailvelope} extensions to steal passwords and PGP private
keys, respectively.

\subsection{Attaching to the browser target}
\label{sec:attaching-to-the-browser-target}

We mentioned in \S\ref{sec:chrome-devtools-protocol}
that extensions using the Debugger API are not supposed nor allowed to attach to the
browser target.
This is presumably to prevent a rogue extension from taking control of an
end-user's browser.
Since they cannot use the \texttt{Target} domain, extensions attach to
targets using the \texttt{chrome.debugger.attach()}
JavaScript binding provided by the Debugger API, which allows an extension
to send CDP commands to, and only to, a given target. That is, if an
extension wants to instrument two different tabs, it will need to attach to
both of them separately.
If it were to exist, a hypothetical extension with access to the browser target
would be capable of instrumenting any tab or extension running in the browser as
the \texttt{Target} CDP domain does not enforce further access control rules.

\vspace{2mm}
\noindent\textbf{Attack vector.}
For reasons similar to those discussed for the Screen Reader extension
(See \S\ref{sec:running-on-webui-tabs}),
there is another special extension that is granted a privileged status
with more capabilities, including the ability to attach to
the browser target using the Debugger API: the Perfetto UI
extension~\cite{cws-perfetto-ui}. This is an official development tool from
Google that interacts with a web app~\cite{perfetto-ui} to profile, record,
and view Chromium execution traces.
To obtain these traces, the Perfetto UI extension uses commands from the
\texttt{Tracing} CDP domain, which are not accessible outside the browser
target.
Unlike the running on WebUI tabs and on other extensions attacks, there
is no flag to replicate this backdoor access. As so, it can only be exploited
by sideloading a clone extension that impersonates the Perfetto UI extension
following the steps we introduced in \S\ref{sec:running-on-webui-tabs}.
This is hardcoded in the Chromium source code
and allows the Perfetto UI extension to attach to the browser target by specifying
the undocumented ``browser'' target ID, as shown in
Listing~\ref{lst:mayAttachToBrowser} and
Listing~\ref{lst:attachToBrowserTarget}.

\begin{lstlisting}[
  float=t!,
  caption={%
    Code to allow attaching to browser target
    (adapted from~\cite{code-debugger-api}).%
  },
  label={lst:mayAttachToBrowser}
]
constexpr char kBrowserTargetId[] = "browser";
constexpr char kPerfettoUIExtensionId[] =
  "lfmkphfpdbjijhpomgecfikhfohaoine";

bool ExtensionMayAttachToBrowser(Extension& ext) {
  return ext.id() == kPerfettoUIExtensionId;
}

bool DebuggerFunction::InitAgentHost() {
  // [...]

  if (
    *debuggee_.target_id == kBrowserTargetId &&
    ExtensionMayAttachToBrowser(*extension())
  ) {
    agent_host_ = CreateForBrowser();
  }

  // [...]
  return true;
}

\end{lstlisting}

\begin{lstlisting}[
  float=t!,
  caption={JavaScript code to attach to the browser target.},
  label={lst:attachToBrowserTarget}
]
const CDP_VERSION = "1.3";
const targetId = "browser";
chrome.debugger.attach({targetId}, CDP_VERSION);
\end{lstlisting}

The browser target that the Perfetto UI extension can attach to is not as powerful
as the actual browser target because some domains and commands are
inaccessible (\eg the \texttt{Network} domain is missing from this CDP host).
Nevertheless, this tool still has access to more domains than it seems to
need to deliver its functionality,
which goes against the principle of least privilege.
One of these domains is the \texttt{Target} domain, which from here is not
capable of communicating with other targets because the
\texttt{Target.sendMessageToTarget} command does not deliver CDP messages to
to their destination.
Another functionality that does not work is the flattened
access~\cite{getting-started-with-cdp} ---an improved mode of operation intended
to deprecate the former command--- as the \texttt{sessionId} property that needs
to be added to CDP messages for this mode to work cannot be sent using the
Debugger API.
However, these limitations can be bypassed to send commands to any
arbitrary target as follows:
\begin{enumerate}
  \item Attach to the browser target using the Debugger API.
  \item Create a \textit{proxy} target (a tab with an \texttt{about:blank}
        URL will do) and attach to it using the
        Debugger API like in the previous step.
  \item Send the \texttt{Target.exposeDevToolsProtocol} command to the
        browser target to expose a pair of JavaScript bindings in the proxy
        target. Then, they will be used to communicate with the desired arbitrary
        target through
        a different communication channel that will allow sending CDP
        messages with the \texttt{sessionId} property.
  \item Use the \texttt{window.cdp} object from the proxy target
	 to attach to the desired final target
        (\eg a WebUI tab or an extension background page) and start
        sending commands to it through the proxy.
\end{enumerate}

\vspace{2mm}
\noindent\textbf{Impact.}
By impersonating the Perfetto UI extension, an attacker can abuse the CDP
browser target to gain all capabilities from the previous attacks combined.
In practice, this means an extension can hijack any regular tab, WebUI page or
extension belonging to any browser context, including incognito windows.

We include two different artifacts to exemplify this attack and its
impact. The first one is
another credit card stealer similar to the one we made for
\S\ref{sec:running-on-webui-tabs}. However, this sample uses the Debugger API
and the browser target to attach to \texttt{chrome://settings/payments}.
The second sample is a browser-wide traffic monitor that logs all requests and
responses from any opened tab or running extension, including incognito windows
and WebUI tabs (see Listing~\ref{lst:attack-6-example} for a simplified code
snippet).

\begin{lstlisting}[
  float=t!,
  caption={Simplified JavaScript code for logging all browser network traffic.},
  label={lst:attack-6-example}
]
// Register CDP event listener
function onEvent(source, method, params) {
  if (params.responseStatusCode) {
    console.log("Response", params.responseHeaders);
  } else {
    console.log("Request", params.request.headers);
  }
}
chrome.debugger.onEvent.addListener(onEvent);

// Attach to browser target and enable domain
const targetId = "browser";
await chrome.debugger.attach({targetId}, "1.3");
await chrome.debugger.sendCommand({targetId},
  "Fetch.enable");
\end{lstlisting}

\section{Discussion}
\label{sec:discussion}

In this section, we discuss the impact of our attacks, our concerns
with the current Chromium extensions architecture, and the effectiveness of
the solutions already implemented or
proposed by the Chromium Development Team.

\subsection{Impact}
The attacks presented in this paper allow a malicious extension to abuse
flaws in the Debugger API to steal sensitive user information and manipulate
runtime behavior (Table~\ref{tab:summary-table}), all while violating basic
security requirements such as isolation or access control
(Table~\ref{tab:requirements-table}).
In addition, since most of the Chromium UI is implemented using WebUI pages
(\S\ref{sec:running-on-webui-tabs}), an attacker is also able to
change browser settings or even experimental flags.
We have also shown how the Debugger API can be misused to modify site
permissions (\S\ref{sec:running-on-regular-tabs}) or bypass security
interstitials without any user awareness or intervention
(\S\ref{sec:running-on-security-interstitial-tabs}).

Our attacks can have a high impact due to the privileged capabilities
an attacker can acquire,
but also because they are rooted in core components of the Chromium Project,
which is critical in today's web browser market.
According to Statcounters~\cite{statcounter-browser-market-share}, Google
Chrome has a 60\% of market share and it is present
in popular forks such as Microsoft Edge, Brave browser, and Opera among
others.
Furthermore, three of our attacks
(\S\ref{sec:running-on-webui-tabs},
\S\ref{sec:running-on-other-extensions},
\S\ref{sec:attaching-to-the-browser-target}) are in part the result of
a questionable approach to integrate two officially-branded Google extensions
with Chromium. This approach consists on hardcoding the IDs of such extensions
into the browser's source code, thus widening the impact of the attacks by
propagating these changes in Chromium forks as well.
Other Chromium-based products may inherit this
artifact and its associated risks, even if said extensions are not intended
to run in them. This might be a even larger problem for
products with a more permissive extension origin policy since they might
not be aware of this threat vector.

\subsection{Root causes}
\label{sec:discussion:root-causes}
The attacks proposed in our work are possible due to a combination of
factors that have shaped core components of the
Chromium project and its development.
We now discuss these enablers.

\vspace{2mm}
\noindent\textbf{Design flaws.}
The attacks described in this paper do not exploit flaws in the
implementation of a specification (\eg memory corruption issues).
Instead, they originate from design decisions where the risk-benefit trade-off
may not be well balanced.
One clear example is the attack described in
\S\ref{sec:running-on-security-interstitial-tabs}. Attaching to security
interstitials using the Debugger API was not possible until Chromium 70
when this capability was intentionally
granted~\cite{gerrit-debugger-interstitials} to prevent the Lighthouse
extension---an official Google development tool---from detaching when running
tests on a web application without offline
support~\cite{gh-lighthouse-interstitials}.
While the implementation of this change did not introduce any bugs \textit{per se},
the design decision behind it is, at a minimum, questionable because security
interstitials are triggered after a critical event that requires user
consent.

The backdoor accesses used in \S\ref{sec:running-on-webui-tabs},
\S\ref{sec:running-on-other-extensions} and
\S\ref{sec:attaching-to-the-browser-target} suggest design
choices that were not sufficiently evaluated from a security standpoint.
These privileges were implemented in such a way that they
blindly trust the ID of an extension without verifying its signature.
In the case of the Screen Reader extension, even the Chromium developer who
introduced the change in October 2011 explicitly acknowledged that this
solution is ``temporary'' and that a long-term alternative is
needed~\cite{chromium-cr-chromevox}.  However, as of the writing of this
paper, this temporary code written 11 years ago has not yet been superseded
nor there is any indication that it will be.

As for the Perfetto UI extension, the change was introduced in January
2020~\cite{gerrit-perfetto-ui-hardcoded} to let
\url{https://ui.perfetto.dev} get traces from all tabs or renderers through
this extension.  Unfortunately, more CDP domains apart from
\texttt{Tracing} were exposed when allowing attaching to a
limited instance of the browser target. It is evident that this decision was not
deeply thought-out and that extensions are not expected to attach to the
browser target given the convoluted workaround we had to come up with for
\S\ref{sec:attaching-to-the-browser-target}.

\vspace{2mm}
\noindent\textbf{Lack of specifications.}
While the Chromium Project has public Design Documents for most of its
components~\cite{chromium-design-documents}, we could not find the
specification for some extension APIs, including the Debugger API. This makes
it challenging to independently
determine whether a particular browser behavior is a feature or a bug.
Thus, we were unsure whether listing and attaching to targets from
incognito windows (\S\ref{sec:listing-active-targets},
\S\ref{sec:running-on-regular-tabs}) was intended or not until April 2022
when we got confirmation from the Chrome Development
Team that it was certainly not intended.
Another example is the \texttt{extensions-on-chrome-urls} flag used in both
\S\ref{sec:running-on-webui-tabs} and \S\ref{sec:running-on-other-extensions}.
While it allows extensions to run on ``chrome://'' URLs,
it also grants the Debugger API access to the ``chrome-extension://'' scheme.
We could not find this behavior documented anywhere.
Once again, we are unsure whether this is intentional or not as the name of
the flag does not suggest it will also apply to browser extensions.
To make matters worse, to this date we still do not know why the flag was
introduced in the first place as the associated Chromium
issue\footnote{See \url{https://crbug.com/174183}.}
has not yet been made public.

\vspace{2mm}
\noindent\textbf{Overpowered APIs.}
Extensions can control almost every single aspect of the browser through the
use of extension APIs. As we demonstrate in this paper,
the Debugger API is one of the more powerful ones, if
not the most.
Some CDP commands can even break basic browser isolation mechanisms
(\S\ref{sec:running-on-regular-tabs}). This allows accessing or modifying any
cookie stored in the browser without the user being clearly aware of
the capabilities of the extension.
Additionally, we have shown how targets from incognito windows are not
isolated either as the Debugger API can list and attach to them even when
the extension has not been granted explicit permission.

\begin{figure}[t!]
  \centering
  \includegraphics[width=0.8\columnwidth]{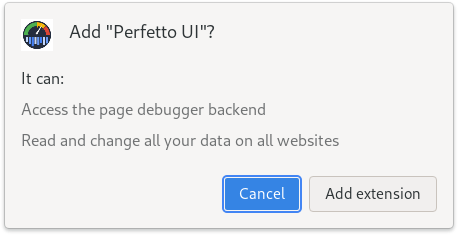}
  \caption{Install confirmation dialog for an extension using the Debugger API.}
  \label{fig:chromium-permissions-warning}
\end{figure}

Despite the Debugger API being a highly privileged and potentially
dangerous capability, the associated risk is not reflected accordingly.
We found all of our attacks violate SR01: when installing an extension that
declares the \texttt{debugger} permission, the install prompt merely shows a vague
warning informing the user that the extension will have access to the
\textit{``page debugger backend''} (Figure~\ref{fig:chromium-permissions-warning}).
We find this risk communication strategy very inadequate.
Other platforms have dealt with apps that abuse powerful permissions in the
past. In the case of Android, this was partially addressed by mandatorily
locking certain resources behind runtime permissions~\cite{android-permissions}.
Although Chromium extensions have a similar concept called ``optional
permissions,'' it is up to the developers to use them or keep requesting a
permission at install-time~\cite{google-permissions}.

Another issue with the Debugger API is that users have little control over
which targets an extension is allowed to debug. As mentioned in
\S\ref{sec:running-on-regular-tabs}, users can force a debugging extension
to detach from all targets by closing an infobar
(\S\ref{sec:cdp-debugger-inforbars}). Nevertheless, extensions can reattach
immediately after.
Even if the user becomes aware of this deceptive behavior, there is little
they can do to prevent it besides uninstalling the malicious extension.
This situation turns the notification infobar into a merely
informative mechanism that can be easily defeated by an effectively
overpowered extension.

\subsection{Solutions}
We reported all of our findings to Google.
The issue related to the attack from \S\ref{sec:listing-active-targets} was
flagged as a duplicate of a similar report\footnote{See \crbug{1236325}.}
by another researcher in August 2021, which stayed stale for months
until we submitted our own report. We did not know this at the time of reporting.
Both this bug and the one we present in
\S\ref{sec:running-on-security-interstitial-tabs}
have already been addressed and fixed by adding appropriate security controls.
The different vulnerabilities that enable the attacks from
\S\ref{sec:running-on-webui-tabs}, \S\ref{sec:running-on-other-extensions} and
part of \S\ref{sec:attaching-to-the-browser-target} were merged into a single
issue\footnote{Reported in \crbug{1276497}.} after Google flagged the rest of
them as duplicates on December 2021. 
More than a year later, this issue is still open and awaiting a fix from the
Chromium Development Team.

The other bug\footnote{Reported in \crbug{1301966}.} that we reported in relation
to \S\ref{sec:attaching-to-the-browser-target} was initially considered a
security vulnerability but then flagged as ``not-a-security-bug,''
consequently making
it public. While the issue remains open and at some point there was some
discussion around preventing sideloaded extensions from attaching to the
browser target, no progress has been made since.

Regardless of whether our attacks exploit actual bugs or intended features,
it is evident that
the Debugger API provides extensions with a wide range of powerful
capabilities, and that such privileged capabilities require extensive and
effective security controls.
At some point during the vulnerability disclosure process, members of the
Chromium Team proposed restricting the Debugger API in its entirety to users
with developer mode enabled.
This was later deemed not ideal as it would impact legitimate developers.
Instead, one strategy we propose for risk reduction would be to follow the
principle of least privilege and redesign a finer grained
permission system for the Debugger API. Examples of this idea abound in
other platforms, including the Linux capabilities introduced in the kernel
starting with version 2.2~\cite{linux-capabilities}, or the splitting of
the \texttt{location} permission in Android into more than one for different
purposes. In the case of the Debugger API, having a separate permission for
each CDP domain will greatly
improve security and reduce the impact of abusive extensions.
In fact, this is explicitly advised in the CRX API Security Checklist from
the Chromium Security Team~\cite{chromium-security-checklist}.

The security principle of Separation of Privilege (\ie granting access based on
meeting multiple independent conditions) can also contribute to risk reduction.
Two easy-to-implement additional
security checks that can be tested when an extension intends to access
the Debugger API are:
\begin{itemize}
  \item\textit{``Trusted'' v. ``non-trusted'' extensions.}
Chromium already can detect the origin of an extension (\eg Chrome Web
Store, sideloaded). The browser could use this information to restrict certain
manifest permissions to ``trusted'' extensions that are signed or coming from a
verified distribution platform.

  \item\textit{Asking for user consent at runtime.}
Instead of just showing a merely informative infobar
(\S\ref{sec:cdp-debugger-inforbars}), we suggest replacing it with a consent
dialog asking the user to grant or deny extension attachment requests.
A similar approach to put sensitive APIs behind a user gesture permission or
a security warning was proposed by a member of the Chromium Development Team.
\end{itemize}

Regarding the Screen Reader and Perfetto UI extensions, hardcoding the
identifier of a privileged subject is not a robust security mechanism and
opens backdoors. Perhaps the safest option would be to convert
them to internal Chromium components. In doing so, the extra capabilities
needed exclusively by these two extensions could be packaged into private
extension APIs accessible to them instead of granting access based on a
hardcoded credential.

To discern whether potential issues
are actual intended features and to ease their assessment,
we suggest including the corresponding technical specification
alongside code changes. This would facilitate to publicly audit code changes
to make sure they comply with the specification. Additionally, having a link between
the design and the implementation makes Design Documents easier to find.

\section{Related work}
\label{sec:related-work}

\vspace{2mm}
\noindent\textbf{Detecting harmful extensions.}
Previous research on the web extensions ecosystem has focused on detecting
malicious browser extensions (MBE). DeKoven \etal describe a methodology
for identifying users visiting Facebook that had MBEs installed in their
browser~\cite{malicious-extensions-at-scale}, which builds upon the work
of Kapravelos \etal in Hulk~\cite{hulk}.
Based on it, they were able
to label close to 2k malicious Chromium and Firefox extensions and notify
their users.  Pantelaios \etal used a sample of almost 1M different
extension releases to propose a system for detecting malicious
updates~\cite{youve-changed}.
Saini \etal introduce attacks where colluding extension share and access
sensitive information without being noticed~\cite{colluding-extensions}.
Similarly, other works focus instead on extensions intentionally leaking
personal data. Chen and Kapravelos developed an automated taint-analysis
framework and found that almost 4k of extensions downloaded from the Chrome
Web Store leaked privacy-sensitive information~\cite{mystique}.
Weissbacher \etal propose Ex-Ray, a technique for dynamically detecting
browsing history leaks through traffic analysis~\cite{exray}.

\vspace{2mm}
\noindent\textbf{Detecting vulnerable extensions.}
Prior studies looked for browser extensions with bugs that
can be exploited by web pages or another extension.
Bandhakavi \etal used static analysis to automatically flag
potentially vulnerable code in legacy Firefox extensions~\cite{vex}.  Fass
\etal improved on this with \textsc{Double}X, a classifier that managed to
accurately detect known flaws on a labeled vulnerable extension
dataset~\cite{doublex}.

\vspace{2mm}
\noindent\textbf{Listing running extensions.}
Sanchez-Rola \etal used timing side-channel attacks to abuse flaws in the
implementation of access control settings in multiple browsers that support
extensions~\cite{extension-resources-control-policies}.
Laperdrix \etal looked at CSS modifications
injected into web pages by extensions to fingerprint the set of extensions
running in a browser~\cite{fingerprinting-in-style}.
Starov and Nikiforakis proposed a similar technique in
XH\textsc{ound}, an automated framework to measures changes in the DOM for
the same purpose~\cite{xhound}.

\vspace{2mm}
\noindent\textbf{Analysis of the browser extensions architecture.}
All the previously mentioned works showcase how powerful extensions
are. However, they mainly focus on identifying abusive or abusable
extensions, not on discussing the browser extension architecture that enables
their behavior.
Reeder \etal performed a user study on warning messages displayed by the
browser~\cite{study-of-user-reactions-to-warnings}. They found that, while
warning design has come a long way, there is still room for improvement as
the context of a warning largely influences the resulting outcome.

\section{Conclusions}
\label{sec:conclusions}

In this work we presented several attacks that exploit vulnerabilities in the
Chromium Debugger API. Our attacks allow a malicious extension to take full
control of the browser, access sensitive information, impersonate other
extensions, and exploit restricted features not intended to be used by any
extension. We demonstrated some of these actions through several PoCs that
we made publicly available. Our attacks affect every major Chromium-based
browser. We reported our findings to Google, which already fixed some of
the vulnerabilities and is addressing the rest at the time of this writing.

Even though inadequate security checks are enablers for some of our
attacks, we believe that the Debugger API grants extensions excessive
capabilities through just one permission, and that the granting
mechanism does not fully convey the risk to the user.
We have provided
constructive discussion on the root cause for most of these vulnerabilities
and potential strategies to improve the security of the Debugger API.

\section*{Acknowledgements}
We are deeply grateful to our anonymous reviewers and to Chromium's Security
Team for their valuable insights and recognition.
This research was supported by
the Spanish Government grant ODIO (PID2019-111429RB-C21 and PID2019-111429RB-C22);
the Region of Madrid, co-financed by European Structural Funds ESF and FEDER
Funds, grant CYNAMON-CM (P2018/TCS-4566);
and by the EU H2020 grant TRUST aWARE (101021377).
José Miguel Moreno was supported by the Spanish Ministry of Science and Innovation
with a FPI Predoctoral Grant (PRE2020-094224).
Narseo Vallina-Rodriguez has been appointed as a Ramon y Cajal Fellow
(RYC2020-030316-I).
The opinions, findings, and conclusions, or recommendations expressed are those
of the authors and do not necessarily reflect the views of any of the funding bodies.

\bibliographystyle{plain}
\bibliography{paper}

\end{document}